\documentclass[doublecol,linenumbers]{epl2} 

\usepackage{amssymb,amsmath,amsfonts}
\usepackage{graphicx,color}
\usepackage[english]{babel}
\usepackage{tabularx}

\usepackage{soul}

\nolinenumbers

\title{Active Colloidal Molecules}
\shorttitle{Title} 

\author{Hartmut L\"owen\inst{1}}

\institute{                    
  \inst{1} Institut f\"{u}r Theoretische Physik II: Weiche Materie, Heinrich-Heine-Universit\"{a}t D\"{u}sseldorf, D-40225 D\"{u}sseldorf, Germany
  }

\abstract{Like ordinary molecules are composed of atoms, colloidal molecules consist 
of several species of colloidal particles tightly bound together.
If one of these components is self-propelled or swimming, novel ``active colloidal molecules" emerge.
Active colloidal molecules exist on various
levels such as ``homonuclear", ``heteronuclear" and ``polymeric" and possess a dynamical function moving as propellers,
spinners or rotors. Self-assembly of such active complexes has been studied a lot
recently and this perspective article summarizes recent progress and gives an outlook to future developments
in the rapidly expanding field of active colloidal molecules. \\ 
\\
\bf{EPL perspective article invited by G. Muga}
}

\begin{document}

\maketitle

\section{Introduction}\label{ra_sec1}

\begin{figure}[] 
		\centering
		\includegraphics[width=0.48\textwidth]{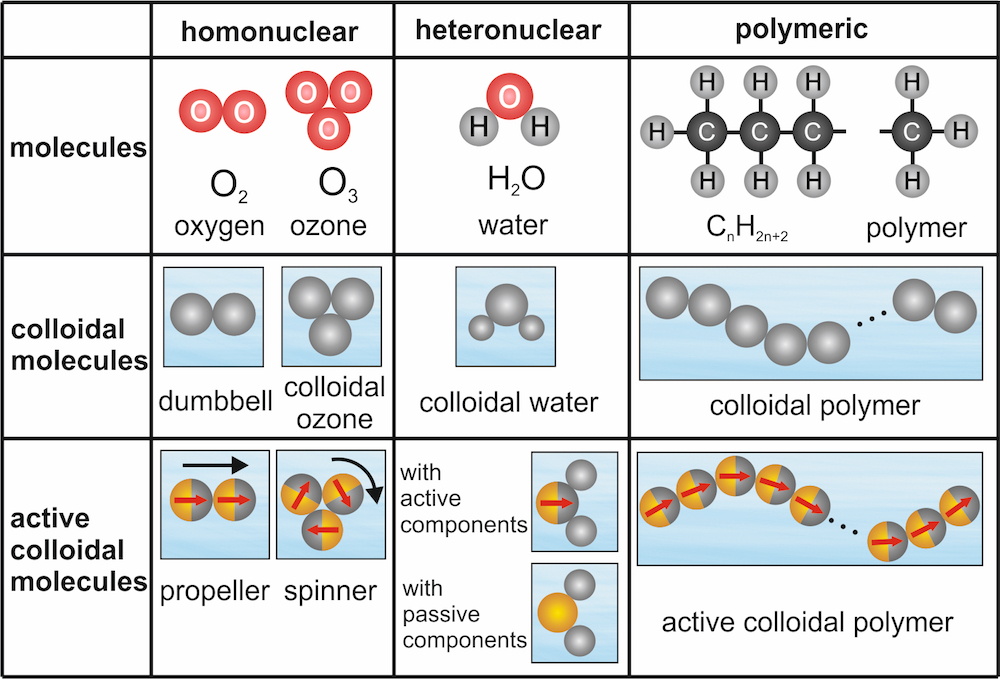}
		\caption{Classification of ordinary and colloidal molecules (schematic). First row:
ordinary homonuclear molecules such as oxygen $O_2$ and  ozone $O_3$, heteronuclear molecules such as
$H_2O$, and polymeric molecules such as a hydrocarbon chain $C_nH_{2n+2}$. Second row:
colloidal molecules composed of colloidal particles (grey spheres) in a solvent (blue background)
including homonuclear dimers and trimers and heteronuclear trimers with two species as well as colloidal polymers.
Third row: active colloidal molecules such as homonuclear dimers (translational ``propeller")
and trimers (rotational ``spinner"). An active entity is indicated by a Janus particles with
the red arrow indicating its self-propulsion direction. Heteronuclear
active colloidal molecules can be divided into two subclasses:
i) composed of mixtures of active and passive building blocks, ii) composed of passive building blocks
which become active when they meet. Active polymeric colloidal molecules consist of self-propelled monomers.}
		\label{fig1}
	\end{figure}

By definition, a {\bf molecule\/} is a group of
two or more atoms held together by chemical bonds. It
is a cornerstone of modern chemistry and physics that matter is composed of such molecules which are
themselves groups of atoms from the periodic table of elements. The periodic
table of elements can therefore be viewed as a ``toolkit`` to construct
and compose molecules and matter in general. 
There are two different
types of molecules: On the one hand, molecules can be composed of identical
atoms referred to as  {\it homonuclear} molecules. An example is
the oxygen molecule $O_2$ or the ozone molecule $O_3$ composed of
oxygen atoms $O$. On the other hand, when molecules are composed of atoms of
different (e.g. two) species they are called  {\it heteronuclear},
an example of which is the water molecule $H_2O$ or an organic hydrocarbon chain
$C_nH_{2n+2}$ serving 
as a standard example for a {\it polymeric} macromolecules, see Figure \ref{fig1} for an illustration.
The typical length scale
for the molecular size amounts to a few Angstroms or nanometers.
Chemical bonding is typically discussed for the case of isolated molecules,
i.e. in the gas phase,
where the surrounding medium is the vacuum.  At room temperature
such a molecular gas can often be described as a classical particle system
such that temperature excites internal translational, vibrational  and rotational modes of the
molecules.\\

The composition principle realized
on the molecular scale can be copied on a larger mesoscopic or ``colloidal"
length scale between a few nanometers and microns. 
In fact, these colloidal particles can be brought together by
attractive effective interaction forces to form a {\bf colloidal molecule}.
This was demonstrated in the beginning of this century by using various
kinds of colloids including patchy particles with designed attraction zones
which form a variety of structures and shapes
similar to ordinary molecules
\cite{Manoharan2003,Blaaderen, Bianchi2006,Glotzer,Kraft1,Kraft2,Wang2012,Kraft_PRE,Shen}.
Like their atomistic counterparts, the colloidal molecules can be classified into {\it homonuclear},
{\it heteronuclear} and {\it polymeric}, see again Figure \ref{fig1}. As for an example of ``homonuclear" and
``heteronuclear" colloidal molecules, trimers
composed of identical spheres (``colloidal ozone molecule") and composed of different
spheres (``colloidal water molecule") can be synthesized; see Ref. \cite{Kraft_PRE} for
visualization of such colloidal molecules. Moreover, there are ``colloidal polymers"
which are chains of linked colloidal
monomer representing realizations of macromolecules in the colloidal domain.
These system have also been synthesized by various means \cite{Schoepe1,Vutukuri,Schoepe3,Schoepe4}.
Mesoscopic colloids offer an important advantage over molecular systems as the
 individual constituents can be designed at wish \cite{micro3dprinting}. Hence a wide range of
colloidal molecules with desired
structural and functionalized properties emerges \cite{Lin} which was an important research line
in the last decades. As an important difference to ordinary molecules,
colloidal molecules are typically suspended in a molecular
carrier fluid indicated as a blue background in Figure \ref{fig1} which exposes them  to
thermal fluctuations induced from the solvent  leading to their Brownian motion.

The key topic of this perspective article is another recent important
step ahead towards {\bf active colloidal molecules}. Active (i.e.\ motile) colloidal molecules are composed of
``active" or ``self-propelled" colloids. The latter are colloidal particles equipped with an internal motor for
their own self-propulsion or, since the particles are embedded in a liquid, for ``swimming".
 Therefore these particles
are sometimes called (synthetic) {\it microswimmers}. By now there are various means to allow colloidal
particles to move on their own, by catalytic reactions, laser-induced diffusiophoresis, external field actuation, etc
see \cite{Elgeti, Gompper2016,our_RMP} for recent reviews. This is highly interesting first 
of all from a basic physics points of view \cite{Ramaswamy2010,Menzel2015,Zottl2016}
as the particles which consume energy are intrinsically in nonequilibrium.
Second, microswimmers are an interesting unanimous realization of moving miroorganisms often leading to
functionalities such as navigation \cite{Dusenbery2009} or self-healing \cite{Trask2007}. 
Third, many applications become possible such as targeted drug delivery via
self-propelled particles \cite{drug_delivery}.

We define {\it active colloidal molecules\/} by a group of colloidal particles with the following conditions:
i) At least one particle is active (or particles get activated during molecule formation, see below), ii) all particles are interacting by attractive body forces even in the
absence of self-propulsion. The second condition implies that collections of particles
bound together by pure hydrodynamic interactions are not termed as ``colloidal molecules". This definition is not
completely sharp. Previously the attractive bond has been postulated to be permanent (like a spring-like force)
\cite{Babel_EPL,Kuechler} but we generalize it here a bit towards other  strong bonding attraction
energies  much larger than the thermal energy $k_BT$. Typically  the bonding force is
then larger than or comparable to forces arising from self-propulsion.
In this context we define an active molecules not as a single swimmer but as an assembly of many swimmers
\cite{Babel_EPL}. Many bead models for a single swimmer as the traditional three-bead
Najafi-Golestanian swimmer \cite{Golestanian_Najafi}
with nonreciprocal dynamics in the relative motion of the beads are in this sense not an active molecule.
In this regard, only
the special case of a heteronuclear molecules with one  passive component or two passive particles
which get active when combined are ambiguous, since such a pair can be simultaneously
considered as a ``single swimmer"
or an ``active molecule".

Figures \ref{fig1} summarizes a possible classification scheme of active colloidal molecules. 
The simplest homonuclear
molecule is composed of two identical self-propelled particles. When combined to a pair by bonding, the dimer propels
translationally along the joint orientation. A trimeric homonuclear ring-like aggregate (also shown in
Figure \ref{fig1}) leads to a spinning molecule or a rotator. The basic difference between ``thermal" colloidal molecules
and active colloidal molecules is their dynamic function: active colloidal molecules move on their own.
This sets them apart for special dynamic functionalities. Clearly different species can be combined in a
next level of complexity leading to heteronuclear molecules \cite{Ebbens2016,Mallory_review}.
In the binary case, two situations can be distinguished:
either there is a combination of active and passive components (e.g. active colloids are mixed with
ordinary passive colloids)
or there is an even more extraordinary case which has no direct analog to ordinary colloidal molecules:
two components which are passive on their own become active when exposed and bound to each other thus
achieving a dynamical function
only when combined \cite{Soto2014,Soto_Golestanian2}. We shall discuss experimental realizations of these
remarkable molecules in the following \cite{Niu2017,Niu2017a,Liebchen}. Finally, polymeric 
active colloidal molecules are conceivable as well; these could be long colloidal chains with
active monomers leading to an active
colloidal polymer or an ``active macromolecule" \cite{Granick_Nat_Materials}.

At this place it is important to distinguish between active colloidal molecules,
for which we assume attractive bonding
between the different constituents,
and purely motility-induced phase separation which nucleates at clusters which occur already
for purely repulsive systems \cite{Buttinoni2013}. We do not discuss these fragile clusters
here in depth since this motility-induced clustering is by now well-established and has been
reviewed already \cite{MIPS_review,Speck_review}. Instead we focus on ``active molecules" where a real
bonding energy (or effective attractive force) 
exists between the ``atoms". However, a strong attractive phoretic attraction 
between neighboring particles in the experimental samples \cite{Theurkauff2012,Palacci2013} 
will bring dynamical clusters to the case considered here.

This perspective article is organized as follows: we first discuss a simple minimal model for
an active colloidal molecule to illustrate its stability and dynamics. We then review more
sophisticated modelling and theoretical predictions obtained mainly by computer simulations
of microswimmer clusters. Then we turn to experimental realizations which have been proposed recently. Finally,
conclusions and perspectives of where this rapidly expanding field is developing to are provided
and some challenges ahead are discussed.

\section{Simplest model of an active colloidal molecule} 

\begin{figure}[] 
		\centering
		\includegraphics[width=0.48\textwidth]{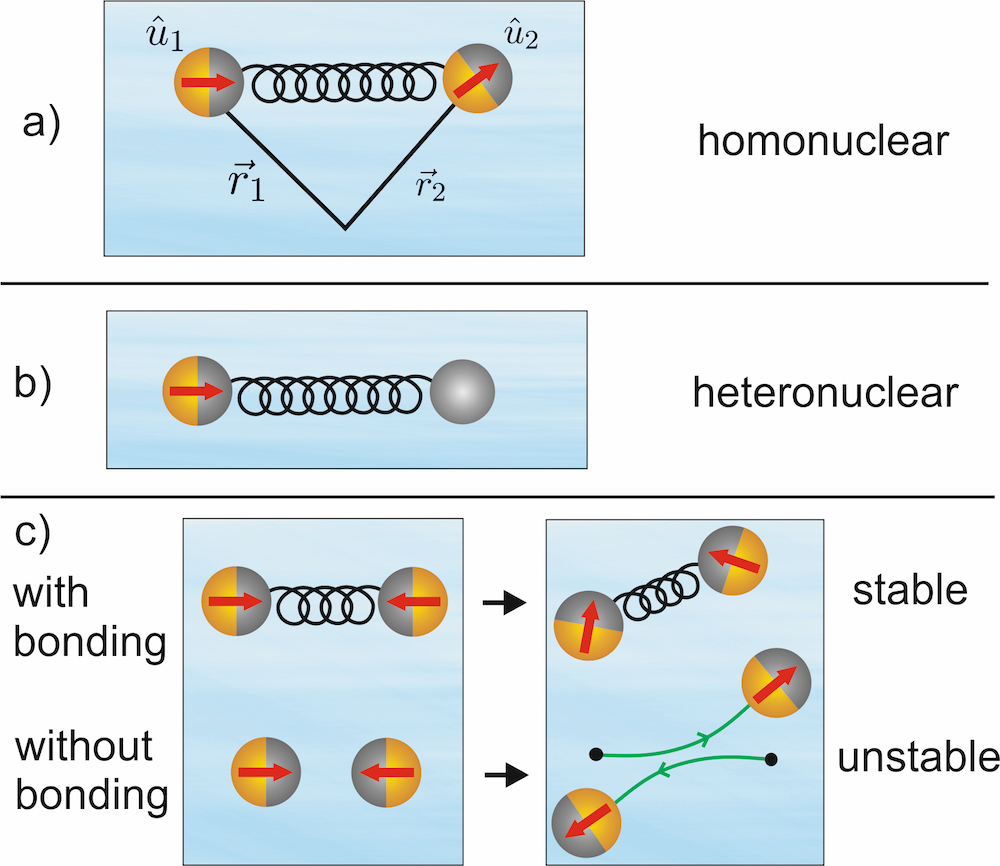}
		\caption{Simple models for an active colloidal molecule. a) Homonuclear active colloidal molecule
composed of two identical self-propelled particles at central positions ${\vec r}_1$ and ${\vec r}_2$
and connected by an effective spring as a ``chemical bond".
The self-propulsion direction (or internal propulsion force) is indicated as a red arrow along the unit vectors
${\hat u}_1$ and ${\hat u}_2$.
b) Heteronuclear active colloidal molecule composed of a self-propelled particle and a passive particle.
c) Homonuclear active colloidal molecule with two opposing self-propulsion forces leading to an overall stable
cluster due to the spring. In the absence of any spring, the pair is long-lived on a time scale set by the inverse 
rotational diffusion constant
but unstable for long time unless further neighbouring particles
accumulate on the pair. The latter situation represents a simple
case for motility-induced clustering in the absence of bonding.}
		\label{fig2}
	\end{figure}

Figure \ref{fig2}a represents the simplest situation of an active homonuclear colloidal molecule, namely
two active particles bound via a spring-like (harmonic) potential, sometimes called an active dumbbell \cite{dumbbell}.
In two spatial dimensions,
the configuration of the two self-propelled particles are described by their  center-of-mass positions
${\vec r}_1$ and ${\vec r}_2$ and their orientational unit vectors ${\hat u}_1 = (\cos \phi_1, \sin \phi_1)$ and
${\hat u}_2 = (\cos \phi_2, \sin \phi_2)$ with $\phi_i$ $(i=1,2)$ denoting the angle of the particle orientation
with respect to the $x$-axis. These unit vectors provide the actual directions of the self-propulsion which is an
inner degree of freedom attached to the particles. Since there is a surrounding  fluid and the particles are micron-sized,
we are typically in the low Reynolds number regime and the simplest description is a completely overdamped dynamics for
the translational and orientational motion ($i=1,2$) neglecting any hydrodynamic interactions mediated by the
flow field of the solvent. Then the equations of motion for the translational and orientational degrees of freedom
read as follows:
\begin{equation} \label{eq:II1}
\gamma \dot{\vec{r}}_i = \gamma v {\hat u}_i - {\partial \over  \partial {\vec r}_i}
V(| {\vec r}_1 -  {\vec r}_2 |)  +     {\vec f}_i(t) 
\end{equation}
and
\begin{equation} \label{eq:II2}
\gamma_R \dot {\phi}_i = g_i(t)   
\end{equation}

Eq.\ (\ref{eq:II1})
can be viewed as a force balance. The term $\gamma \dot{\vec{r}}_i$ on the left hand side is the Stokes drag force
with $\gamma$ denoting the Stokes drag translational friction coefficient. The term  $\gamma v {\hat u}_i$
represents the self-propulsion which couples translation and rotation in  a nontrivial way
where $v$ is the imposed self-propulsion speed. The spring interaction force $-{\partial \over  \partial {\vec r}_i}
V(| {\vec r}_1 -  {\vec r}_2 |)$ is derived from a harmonic potential

\begin{equation} \label{eq:II3}
V(r)= {k\over 2} ( r- \ell)^2    
\end{equation}
of a spring with finite rest length $\ell$ and spring constant $k$.
Clearly the spring interaction force is reciprocal, i.e. actio is minus reactio,
such that Newton's third law is fulfilled.
Finally, ${\vec f}_i(t)$ are Gaussian random forces
(``noise") stemming from the solvent kicks with zero mean and a variance between two Cartesian components $\alpha$ and $\beta$
of $<f_i^{\alpha}(t) f_j^{\beta}(t')> = {2{k_BT}{\gamma}}\delta_{\alpha\beta}\delta_{ij}\delta(t-t')$ where
$T$ denotes some effective noise strength
or effective temperature. Note that $T$ does not necessarily correspond to the real temperature as we are far
from equilibrium.
Likewise Eq.\ (\ref{eq:II2}) represents a torque balance between Stokes drag torque $\gamma_R \dot {\phi}_i$ on the left hand side
and a random Gaussian torque $g_i(t)$ with zero mean and variance
$<g_i(t) g_j(t')> = {2{k_BT}{\gamma_R}}\delta_{ij}\delta(t-t')$. Here $\gamma_R$ is the rotational friction coefficient.

It is instructive to consider a configuration for a homonuclear pair where the two particles have opposing
propulsion directions as shown in Figure \ref{fig2}c. In fact, if the two self-propulsion forces are exactly opposing
and are not too large (i.e. $v > k\ell/\gamma$),
there is a distance where they are exactly compensated by the repulsive spring forces. Therefore, in the absence of noise,
a pair will not move in this special configuration. This static configuration can be  characterized
by a six-component multivector
$({\vec r}_1^*, {\vec r}_2^*, {\phi}_1^*=\pi, \phi_2^*=0)$ summarizing the positions and
orientations of the two particles.
Expanding the equations of motion around this static configuration  to
linear order in the six difference variables $X =({\vec r}_1(t) - {\vec r}_1^*$, ${\vec r}_2(t) - {\vec r}_2^*$,
${\phi}_1 - {\phi}_1^*$, ${\phi}_2 - {\phi}_2^*)$ yields a matrix equation of the form

\begin{equation} \label{eq:II4}
{\dot X} = \bar{\bar{A}} \cdot X   
\end{equation}
where $\bar{\bar{A}} $ is a $6\times 6$ ``dynamical" matrix which contains real and constant matrix elements.
The corresponding eigenmodes  around the static
reference point $X=0$ are obtained by plugging an ansatz into Eq.\ (\ref{eq:II4}) of the form

\begin{equation} \label{eq:II5}
X(t) =  X_{\lambda}(0) \exp (\lambda t)   
\end{equation}
where the real part of $\lambda$ is the growth rate and $X_{\lambda}(0)$ the associated eigenvector.
If a solution with a positive real part of  $\lambda$ is possible, the configuration is unstable with respect to the noise,
while a negative $\lambda$ means a
damped eigenmode that is stable against the noise. For the intermediate vanishing case, $\lambda =0$,
the situation is marginal or neutral; this often reflects a continuous symmetry
in the system which leads to a zero eigenmode.

Let us first discuss the simplest reference case where the spring is applied and has a finite extension($\ell>0$, $k>0$) but
the self-propulsion force is vanishing ($v=0$). In this case all forces are reciprocal, the $6\times 6$ matrix
$\bar{\bar{A}} $ is symmetric and can be diagonalized. One eigenvalue is negative and the remaining
five eigenvalues are zero. The latter correspond to five different symmetry operations such as:
i) global translation in $x$-direction, ii) global translation in $y$-direction, iii) global rotation of the whole system,
iv) individual internal rotation of the left particle, v) individual internal rotation of the right particle.
The eigenmode belonging to the negative eigenvalue corresponds to a
radial distortion of the particle pair, this is kept stable against the noise via the attractive bonding.

If the self-propulsion is finite ($v>0$), the matrix is not any longer symmetric, this is a consequence of the fact
that self-propulsion leads to nonreciprocal interactions \cite{Soto2014,Soto_Golestanian2,Ivlev2015}. Nevertheless
there is still one negative eigenvalue which
again corresponds to the radial distortion kept stable via the harmonic bonding.
But there are not any longer 5 clean zero eigenvalues.
Remember that not any real unsymmetric matrix can be diagonalized (even not in the complex), it can only
be brought to the Jordan normal form in a suitable basis. This is a basic difference to passive colloidal molecules
where the dynamical matrix is symmetric and diagonalizable.
In particular, in the special
case of vanishing spring constant $k$ (see Figure \ref{fig2}c again),
there are no stable nor unstable modes but neutral or marginal ones. This configuration is not stable:
following weak noise the particles will scatter around each other and move away from each other, 
see the possible trajectories for
self-propelled particles in Figure \ref{fig2}c.
This set-up corresponds to pure motility-induced clustering without any bonding pointing to the finite life of any
cluster for self-propelled particles with repulsive interactions, see also possible trajectories for
self-propelled particles in Figure \ref{fig2}c.

A heteronuclear situation shown in Figure \ref{fig2}b can be described by similar
overdamped equations of motion (\ref{eq:II1})-(\ref{eq:II2}) \cite{Kuechler}
except for the fact that rotational motion is irrelevant for a passive sphere. The stability analysis is similar in
spirit than for the homonuclear pair.

The simple example shows two basic facts: i) attractive bonding is needed for stable clusters
ii) the motion of a pair or a  general molecule is more complicated than that of the individual
self-propelled particles.

\section{Theory and simulation of active colloidal molecules} 

Let us now discuss recent models for active colloidal molecules which were explored by theory and computer simulations.
Experimental realizations are subsequently mentioned in the next section.

\subsection{Spring-like interaction forces} 

Harmonic springs (see Eq.\ (\ref{eq:II3})) constitute the simplest models for bonding. In particular these forces are obtained when
expanding the equations of motion around  a stable configuration, see again the discussion in the previous section.
Heteronuclear active colloidal molecules where one active particles is linked to passive spheres via springs were
proposed in Ref.\cite{Kuechler}, see also \cite{Misbah}. The resulting active colloidal molecules
can alternatively be viewed as a deformable
swimmer. The importance of deformability was emphasized when the swimmer is exposed to external vortical flow fields
\cite{Kuechler,vortical}.
It turns out that the degree of deformability decides whether the swimmer is
getting drowned in a swirl or not \cite{Kuechler}.

If several active particles are connected by springs along one dimension we arrive to the polymeric
case of  an active  polymer chain, see again Figure 1. The first model for active polymers
was proposed by Kaiser et al
\cite{Kaiser_active_polymer}. It is the natural extension of the harmonic model discussed
for two particles in the previous section to a long chain. In this sense it corresponds
to the active extension of the traditional Rouse model \cite{DoiEdwards_book,Kaiser_active_polymer,Rabin}.
As a result, it was found \cite{Kaiser_active_polymer}
that the polymer swelling is significantly affected by activity. Subsequently, other
models for an active polymers were proposed and discussed such as active worms \cite{Elgeti_worms} and active
filaments \cite{Laskar,Henkes} and polymers with non-thermal active noise \cite{Eisenstecken1,Eisenstecken2}
mimicking the internal activity of the monomers. For more details we refer to  the
recent review in Ref.\cite{review_active_polymers}.

Finally, a  full two-dimensional network of springs connecting active particles to a
solid sheet has been proposed and analyzed by Ferrante et al
\cite{Huepe} revealing new dynamical modes. This can be considered as a limit of a flat
active colloidal macromolecule, analogous to an active molecular graphene sheet.

\subsection{Active colloidal molecules governed by dipolar interactions} 

The interactions between two dipoles depends on their relative orientation and position,
it can be both attractive and repulsive. The attractive case leads to a tendency of chaining in dipolar suspensions.
Recently clusters of active colloidal particles kept together by (e.g. magnetic)
dipolar attractions have been realized \cite{Gemming} scales. These are an ideal set-up for magnetic active molecules.
The classical ground-state minimizing the total interaction energy of $N$ hard-sphere dipoles with a central permanent
dipole moment  is a linear chain for $2\leq N\leq3$ and a ring for $4\leq N \leq 13$ \cite{Messina}. Adding
self-propulsion to these molecules induces novel effects. These have first been explored 
in the absence of hydrodynamic interactions \cite{Popowa}.
As a result, a linear chain propagates and a ring rotates
but they are stable for any self-propulsion speed $v$. Starting with an initial metastable $Y$-like
configuration results in seven different steady states including a {\it fission} of the initial $Y$-junction
\cite{Popowa}. This demonstrates that self-propulsion has subtle effect on molecules formation and structure.

Subsequently the effects of hydrodynamic interactions mediated by the solvent
on the motion of dipolar clusters was explored. First of all, a straight trimer of hard sphere dipoles
with permanent harmonic bonds was considered in Ref.\  \cite{Babel_EPL}. With increasing self-propulsion,
an oscillatory instability was observed which  is accompanied by a
corkscrew-like swimming trajectory of increasing radius. Hydrodynamic interactions in active magnetic clusters
were studied as well by Guzman-Lastra et al
\cite{Guzman-Lastra}. It was shown that fission and fusion processes of colliding clusters
may occur resulting from a competition
between dipolar and magnetic interactions. For instance, a linear dipolar chain, which would be absolutely stable
in the case of neglected hydrodynamic interactions \cite{Popowa},  will break into segments if
the particle swim as ``pushers" but remains stable if they are ``pullers" \cite{Guzman-Lastra}.

\subsection{Attractive chemotaxis} 

Many microorganisms but also synthetic colloids sense a chemical which is emitted
by other particles and adjust their motion towards the positive or negative gradient of the chemical concentration field
a phenomenon which is called {\it chemotaxis}. It is an old idea of Tsori and de Gennes
\cite{Tsori} that particles which are coupled chemotactically are equivalent to a classical Coulomb system \cite{Benno_chapter}.
The underlying reason for the ``Coulomb analogy" is that the stationary solution for the diffusion equation
with a constant ejection source is an orbital around the secreting particle which decays as the inverse distance,
formally exactly the same behaviour as for  the classical Poisson equation of electrostatics
\cite{Benno_chapter}. If in a two-component system,
one species is attracted towards the chemical which is secreted by the other species (chemo-attraction),
chemotaxis is formally identical to a Coulomb
interaction although there is more freedom as the interaction forces are not necessarily reciprocal
\cite{Soto_Golestanian1,Soto_Golestanian2,Liebchen,Bartnick_JPCM}. This provides an enormous
possibility to construct active colloidal molecules with a dynamic function such as
movers, rotators and circle swimmers. In the opposite case of chemo-repulsion there is no bonding
and hence no colloidal molecules emerge.

In this context, it is important to note that the particle of each species on their own are passive, i.e. they would not
move in the absence of particles of the other species. Hence passive building blocks merge and spontaneous acquire
activity by merging. This is what is shown in the third row of Figure \ref{fig1}.

\subsection{Other attractive interactions} 

Other types of attractive interaction for microswimmer bonding have been proposed.
In Ref. \cite{Mallory/CacciutoPRE2016} clusters arise from a lock-and-key shape of particles
like triangular shapes with an opening angle of $2\pi/n$ with $n>1$ an integer number.
These cluster are stabilized by additional attractions and may thereby considered to be active colloidal molecules.
Cluster bound via attractive patches were also proposed and studied \cite{Mallory_NJP} driven by the general
idea to use patchy colloids as active systems. Also swimmers interacting with the traditional Lennard-Jones pair potential
were studied in order to access their nonequilibrium critical behaviour \cite{Prymidis}. For strong enough bonding
these clusters are pretty stable exhibiting active colloidal molecules.

Finally recently binary passive building blocks were studied which when merged get active, such as two individual hemispheres
of a split Janus particle. The attraction was modelled close to an experimental realization to be an attractive Casimir force
appropriate as an effective interaction between the passive building blocks when immersed in a near-critical solvent
\cite{Liebchen}. In this model a variety of active colloidal molecules was discovered, too.

\section{Experimental realization of active colloidal molecules} 

\begin{figure}[] 
		\centering
		\includegraphics[width=0.48\textwidth]{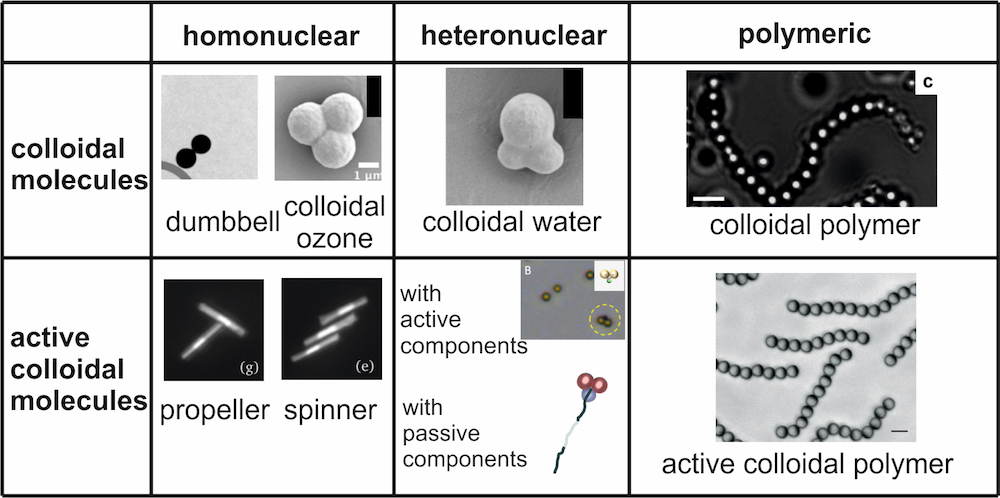}
		\caption{Same as Figure 1, but now not schematic but with actual realizations of active colloidal molecules.
 First row:
colloidal molecules composed of colloidal particles
including homonuclear dimers (image from Ref.\ \cite{Demiroers}) and trimers
(image from Ref.\ \cite{Kraft_PRE}) and heteronuclear trimers  (image from Ref.\ \cite{Kraft_PRE})
with two species as well as a colloidal polymer  (image from Ref.\ \cite{Vutukuri}).
Second row: active colloidal molecules such as homonuclear dimers (translational ``propeller")
(image from Ref.\ \cite{Wykes})
and trimers (image rotational ``spinner")  (image from Ref.\ \cite{Wykes}).  Heteronuclear
active colloidal molecules:
i) composed of mixtures of active and passive building blocks
(image from Ref.\ \cite{Guan}), ii) composed of passive building blocks
which become active when they meet  (image from Ref.\ \cite{Liebchen}).
Active polymeric colloidal molecules   (image from Ref.\ \cite{Granick_Nat_Materials}).}
		\label{fig4}
	\end{figure}

The wonderful world of active colloidal molecules is not pure fiction but these artificial
complexes have meanwhile been largely synthesized and realized. It is instructive to consider the
classification as shown in Figure \ref{fig1} again and draw actual micro-images from recent experiments instead of
the schematic images. This is summarized in Figure \ref{fig4}. In fact {\ all} the different colloidal
molecules shown there have been fabricated and discovered. In what follows we review some basic progress in these experiments.

First of all a harmonic or spring-like bonding can be achieved by attaching DNA bundles onto 
active colloids as recently demonstrated \cite{Dreyfus,Schoepe4}. This was
shown for a chain of passive particles which can be activated by an external magnetic field \cite{Dreyfus}.
Moreover a DNA bundle was used as a flagellum
driving a single colloidal sphere \cite{Maier} but it can likewise also be used to bind two active spheres permanently. A
dimer composed of two active Janus particles was studied as the simplest {\it homonuclear\/} colloidal molecule
\cite{Ebbens_bi,Lammert}. Homonuclear molecules consisting of two to four metallic rods
were found to exhibit interesting structures such as a $T$-shape in Ref. \cite{Wykes}, see also Figure \ref{fig4}.
Electric-field controlled structures of active molecules including {\it active colloidal
polymers} were proposed in Refs.\ \cite{Granick_Nat_Materials,Vutukuri2}.
Another homonuclear ``polymer" consisting of four cubic monomers was found in Ref.\ \cite{Cheng}.
{\it Heteronuclear\/} assemblies of active and passive colloids were
observed for catalytically powered \cite{Sen_2013,Sen_2015}, light-activated
\cite{Peer_Fischer,Guan} and electric-field controlled \cite{Granick_Angewandte} self-propulsion mechanisms.
Finally,  a rich self-organization kit for constructing tight
colloidal molecules on various levels using sequential capillary assembly was provided \cite{Isa}.

Second, there is a broad variety by now for {\it magnetic dipolar swimmers} self-assembled to active molecules. These
can be chains of propellers or rotators \cite{Tierno,Tierno2} as well as  elongated
magnetic particle clusters  \cite{Gemming}
or large colloidal asters  \cite{Aranson1} energized by an alternating magnetic field
\cite{Aranson2}.

Finally there are several examples for {\it passive entities which form an active molecule when combined}.
The first example is a so called modular swimmer consisting of an ion-exchange resin and passive colloids
as proposed by Niu et al \cite{Niu2017,Niu2017a}.  The flux of ions 
towards the resin causes flow which advects passive colloids resulting in a moving composite object
which is a colloidal molecule. For a pure suspension of ion exchangers either one-component or two-component
(cationic and anionic) exchangers colloidal molecules were also obtained \cite{Niu2017,Niu2017a}. 
Finally two kind of passive spherical beads interacting via attractive Casimir forces were studied by Schmidt et al
\cite{Liebchen} which become active under light illumination when merged. This provides an important control
of active molecules and their formation via external light.

\section{Conclusions and challenges ahead} 

In conclusion we have outlined a novel route towards active complexes as ``active colloidal molecules".
Like ordinary molecules are composed of atoms, the same idea can be applied to the colloidal mesoscale
by creating new active composite aggregates merging single active particles.
The self-organization into stable active clusters which possess characteristic dynamical and structural functions has
revealed new nonequilibrium physics and is promising for many applications such as functionalized nanomachines.

In the field of active colloidal molecules, the following challenges are lying ahead:

i)  Once single molecules are established, a possible next step is to mix them to obtain more complex colloidal
structures. These can be based on a lock-and-key principle, on other complex patchy bonding or on hydrodynamic
lubrication \cite{Liverpool}.

ii) Again, once single molecules are established, the next step is go to finite concentration and condensed phases.
One will expect a complex ``zoo" of
different living structures for strongly interacting systems. Few examples include: motility-induced clustering
for active colloidal molecules, active crystals of colloidal molecules which are expected to have a rich  dynamics and a complex melting scenario \cite{Speck,Menzel_PRL,Gonnella,Dauchot}.

iii) Another  next natural step to go is to move from few-particle molecules to ``colloidal macromolecules"
with new hierarchial structures on a supramolecular scale. There is an almost unlimited potential to combine
new building blocks on any scale to arrive at complex particles which can posses interesting structures.
One example are hollow particles which push solvent through the holes as recently realized for active sperm
attached to a hollow tube, so-called spermbots \cite{Oliver_Schmidt,Magdanz}.

iv) So far we have discussed active colloidal molecules in a Newtonian fluid.
New physics arises if the background medium is changed towards a complex fluid which can be viscoelastic
(see e.g.\ \cite{viscoelastic}),
or if the background itself is another kind of soft matter.

v) In general active colloidal molecules will challenge existing theories for nonequilibrium systems
such as dynamical density functional theory (see e g.\ \cite{Saha})  or mode coupling
theory (see e.g.\ \cite{MCT1,MCT2}). These theoretical approaches  need to be generalized
or completely newly founded if activity comes into play.

\acknowledgments
I thank Liesbeth M. C. Janssen, Avni Jain, Andreas M. Menzel and Benno Liebchen for helpful discussions.
H.L. gratefully acknowledges support by the Deutsche
Forschungsgemeinschaft (DFG) through grant
LO 418/19-1.

\bibliographystyle{eplbib}
\bibliography{eplbibfile}

\end{document}